\documentclass[conference]{IEEEtran}
\ifCLASSINFOpdf
\else
\fi
\usepackage{cite}
\usepackage{xcolor}
\usepackage{graphicx}
\usepackage{amsmath}
\usepackage{amsfonts}
\usepackage{physics}
\usepackage{algorithm}
\usepackage[noend]{algpseudocode}
\usepackage{tabularx}
\usepackage{hhline}

\makeatletter
\def\BState{\State\hskip-\ALG@thistlm}
\makeatother

\makeatletter
\newsavebox\myboxA
\newsavebox\myboxB
\newlength\mylenA

\newcommand*\xoverline[2][0.75]{%
    \sbox{\myboxA}{$\m@th#2$}%
    \setbox\myboxB\null
    \ht\myboxB=\ht\myboxA%
    \dp\myboxB=\dp\myboxA%
    \wd\myboxB=#1\wd\myboxA
    \sbox\myboxB{$\m@th\overline{\copy\myboxB}$}
    \setlength\mylenA{\the\wd\myboxA}
    \addtolength\mylenA{-\the\wd\myboxB}%
    \ifdim\wd\myboxB<\wd\myboxA%
       \rlap{\hskip 0.5\mylenA\usebox\myboxB}{\usebox\myboxA}%
    \else
        \hskip -0.5\mylenA\rlap{\usebox\myboxA}{\hskip 0.5\mylenA\usebox\myboxB}%
    \fi}
\makeatother

\begin{document}
%
\title{Multi-Pose Fusion for Sparse-View CT Reconstruction Using Consensus Equilibrium}


\author{
\IEEEauthorblockN{Diyu Yang}
\IEEEauthorblockA{\textit{School of Electrical and Computer Engineering}\\
\textit{Purdue University}\\
West Lafayette, IN, USA\\
yang1467@purdue.edu}\\   
\IEEEauthorblockN{Gregery T. Buzzard}
\IEEEauthorblockA{\textit{Department of Mathematics}\\
\textit{Purdue University}\\
West Lafayette, IN, USA\\
buzzard@purdue.edu}
\and
\IEEEauthorblockN{Craig A. J. Kemp}
\IEEEauthorblockA{\textit{Eli Lilly and Company}\\
Indianapolis, IN, USA\\
kemp\_craig\_a@lilly.com}\\[0.4cm]  
\IEEEauthorblockN{Charles A. Bouman}
\IEEEauthorblockA{\textit{School of Electrical and Computer Engineering}\\
\textit{Purdue University}\\
West Lafayette, IN, USA\\
bouman@purdue.edu}
}


%


\maketitle

\begin{abstract}
CT imaging works by reconstructing an object of interest from a collection of projections. Traditional methods such as filtered-back projection (FBP) work on projection images acquired around a fixed rotation axis. However, for some CT problems, it is desirable to perform a joint reconstruction from projection data acquired from multiple rotation axes.

In this paper, we present \emph{Multi-Pose Fusion}, a novel algorithm that performs a joint tomographic reconstruction from CT scans acquired from multiple poses of a single object, where each pose has a distinct rotation axis. Our approach uses multi-agent consensus equilibrium (MACE), an extension of plug-and-play, as a framework for integrating projection data from different poses. We apply our method on simulated data and demonstrate that Multi-Pose Fusion can achieve a better reconstruction result than single pose reconstruction.

\end{abstract}


%
\IEEEpeerreviewmaketitle

\section{Introduction}

In computed tomography (CT), the projection data is usually acquired from a fixed pose of the object, where different views are measured around a fixed rotation axis. 
However, for some CT applications, it may be more desirable to collect multiple sets of CT scans taken from different poses of the same object. 
For example, a common type of artifact in clinical CT imaging is metal artifacts, and a popular way of reducing metal artifacts is to acquire one or more tilted CT reconstructions from different angles of the same object \cite{MARsummary,Brown1999,Lewis2010,luckow2011tilting, ballhausen2014post,kim2019additional,branco2020development}. 
Therefore, when multiple sets of projection data exists, it is desirable to form a joint reconstruction from all projection data. We call this problem the multi-pose reconstruction problem.

Model-based iterative reconstruction (MBIR) \cite{MBIR} has become a popular method for tomographic reconstruction in scientific and industrial applications when the data is limited or noisy \cite{thibault2007three,zhang2013model,jin2015model,kisner2012model}.
In particular, MBIR and the more general Plug-and-Play (PnP) methods \cite{PnPPrior, PnPPriorSuhas} have been shown to be superior to more traditional direct reconstruction methods such as FDK when the view sampling is very sparse \cite{ThiloPaper}.
However, naive implementation of multi-pose MBIR reconstruction is impractical because it would require highly complex and specialized software that jointly incorporates the system matrices for each distinct pose.

The PnP framework was first proposed as a method for modeling the prior distribution of an image with a denoiser  \cite{PnPPrior,PnPPriorSuhas}.
However, more recently Multi-Agent Consensus Equilibrium (MACE) \cite{CE} has been proposed as a generalization of PnP in which multiple agents can be used to characterize different objectives in an inverse problem \cite{CE,MACE4D2019,MACE4D,FCI2022}.
An important advantage of MACE is that it is based on computing the solution to an equilibrium problem, so it can be used to solve inverse problems when there is no easily defined cost function to minimize.

\begin{figure}[t]
\centering
   \includegraphics[width=\linewidth]{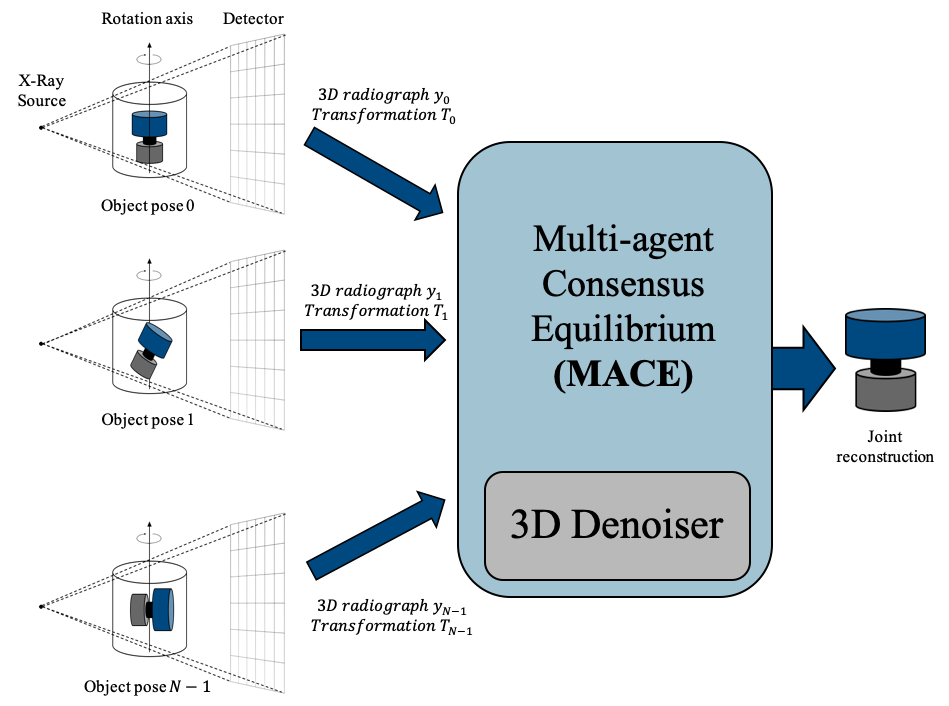}
   \caption{Multi-Pose Fusion Overview: Multiple sets of projections are taken from different poses of the object. The projection data are then fused by MACE framework to form a joint reconstruction.}
   \label{fig:MPF-overview}
\end{figure}

In this paper, we introduce \emph{Multi-Pose Fusion}, a novel CT reconstruction algorithm that fuses measurements from multiple poses of the object and performs a joint reconstruction. 
Figure~\ref{fig:MPF-overview} illustrates the basic concept of the Multi-Pose Fusion algorithm. 
Multi-Pose Fusion performs a reconstruction by using MACE to fuse the data from multiple poses into a single reconstruction.
The MACE algorithm is implemented using a set of agents, where each agent corresponds to the update for a single pose of the object.
MACE then defines a precise criterion for the solution along with an algorithm for computing the solution.
Perhaps most importantly, the resulting algorithm has a simple modular implementation using standard CT reconstruction software.
We compare Multi-Pose Fusion (MPF) results to single pose MBIR and PnP-based reconstructions on a simulated data set and demonstrate that it achieves better reconstruction quality with reduced artifacts and noise, and improved detail.

\section{Problem Formulation}

In multi-pose CT imaging, multiple sets of CT scans are taken using different poses of the object as illustrated in Figure~\ref{fig:MPF-overview}. 
Notice that the imaging geometry is the same for each pose.
However, in practice some poses provide more useful information, particularly when the object contains dense or even opaque components that may obscure portions of the object.
The objective of multi-pose CT reconstruction is then to perform a joint MBIR reconstruction from scans acquired from multiple poses of the object. 

Let $K$ be the number of measurement poses, $M_k$ be the number of CT measurements obtained from pose $k$, and $N$ be the number of voxels of the object to be recovered. 
We define $y_k \in \mathbb{R}^{M_k}$ to be the sinogram measurements at pose $k$, where $k\in\left\{0,...,K-1\right\}$, and we define $x\in\mathbb{R}^{N}$ as the image vector containing attenuation coefficients to be recovered in the reconstruction coordinate system.

For each pose $k$, we also define a transformation function $x_k = T_k x$ where $x$ is the object represented in the common reconstruction coordinate system and $x_k$ is the object represented in the $k^{th}$ pose.
So intuitively, $T_k$ transforms the object from the common reconstruction coordinate system to the posed coordinate system.
In practice, $T_k$ typically implements a rigid body transformation \cite{rigidtransform}, so it requires that the discretized function be resampled on the transformed sampling grid. 
This process requires some form of spline-based interpolation algorithm \cite{de1978practical}.
We will also require an approximate inverse transformation $T_k^{-1}$. Since both transforms require interpolation and resampling, we note that they will not in general be exact inverses of each other.

Using this notation, the forward model at each pose can then be expressed as 
\begin{equation}
    y_k = A_k T_k x + w \ ,
\end{equation}
where $A_k \in \mathbb{R}^{N\times M_k}$ is the scanner system matrix for the $k^{th}$ pose and $w\sim N(0, \alpha \Lambda_k^{-1})$ is additive noise.

The joint MBIR reconstruction for the multi-pose problem is then given by
\begin{equation}
    \label{eq:inverse_problem}
    \hat{x}_{MAP} = \arg\min_{x}\left\{ \sum_{k=0}^{K-1} f_k(x)+h(x)\right\}
\end{equation}
with data fidelity terms given by $f_k (x) = -\log p(y_k |x) + \mbox{const}$ where
\begin{equation}
    \label{eq:sum_of_pose_costs}
    f_k (x) = \frac{1}{2} \lVert  y_k-A_kT_k x  \rVert_{\Lambda_k}^2
\end{equation}
and a prior term given by $h(x)=-\log p(x)$ that imposes regularity.

Notice that direct implementation of \eqref{eq:inverse_problem} is difficult since it requires that software be written to minimize a sum of complex tomographic reconstruction terms each with a different transformation $T_k$.

Alternatively, one can compute the MBIR reconstruction by using consensus ADMM \cite{BoydADMM} to minimize the sum of $K+1$ terms consisting of $h$ and the $K$ terms in \eqref{eq:sum_of_pose_costs}.
However, this approach has a number of serious disadvantages.
First, the proximal map terms required for each pose will be very computationally expensive to compute. 
Second, we can improve reconstruction quality by replacing the prior term $h(x)$ with a PnP denoiser.

\section{MACE Formulation of Multi-Pose Reconstruction}

In this section, we introduce the MACE framework for solving the multi-pose reconstruction problem \cite{CE,FCI2022}.

To do this, we first define $x^\prime = F_k (x)$ to be an agent for the $k^{th}$ pose.
Intuitively, the function of the agent is to take a reconstruction $x$ and return a reconstruction $x^\prime$ that better fits the measurements $y_k$ associated with the data from pose $k$.
Ideally, one might choose to use the following proximal map as an agent.
\begin{eqnarray}
\label{CGPM}
\tilde{F}_k (v) \!\!\!\!\!\!
    &=&\!\!\!\! \arg\min_{x}\left\{ f_k (x) + \frac{1}{2\sigma^2}\lVert x-v\rVert^2\right\} \\
    &=&\!\!\!\! \arg\min_{x}\left\{ \frac{1}{2} \lVert y-A_k T_k x \rVert_{\Lambda_k}^2 + \frac{1}{2\sigma^2}\lVert x-v\rVert^2\right\}
\end{eqnarray}
However, this agent is computationally expensive and difficult to compute since it requires that the transformation $T_k$ be integrated into the reconstruction software.

Alternatively, we propose to use a \emph{Conjugate Proximal Map} as our agent given by
\begin{equation}
    \label{eq:CGPM_final}
    F_k (v) = T_k^{-1} F (T_k v ; y_k ) \ ,
\end{equation}
where $F(v; y)$ is the standard proximal map in reconstruction coordinates given by
\begin{equation}
    \label{eq:standard_prox}
    F(v; y ) = \arg\min_{x} \left\{ \frac{1}{2} \lVert y-A_k x \rVert_{\Lambda_k}^2 + \frac{1}{2\sigma^2}\lVert x - v\rVert^2\right\}.
\end{equation}
Notice that the conjugate proximal map of \eqref{eq:CGPM_final} can be computed easily since it requires only the computation of the standard proximal map of \eqref{eq:standard_prox} in the standard coordinates and pre- and post-composition with the spline-based maps $T_k$ and $T_k^{-1}$.
In fact, software for computing the proximal map in \eqref{eq:standard_prox} is openly available \cite{svmbir, mbircone}.

For the prior model, we will use a variation of the PnP prior known as multi-slice fusion \cite{MACE4D}.
The multi-slice fusion uses three denoising operators each applied along a different set of 2D slices corresponding to $(x,y)$, $(x,z)$ and $(y,z)$ coordinates.
We denote these three denoising agents by $F_K$, $F_{K+1}$, and $F_{K+2}$. 

For notational simplicity, we define the stacked set of agents as
\begin{equation}
\label{eq:stacked_agents}
{\bf F} ( {\bf w})= \left[ F_0(w_0), \cdots , F_{K+2} ( w_{k+2} ) \right]
\end{equation}
where $\bf w$ is the stacked input vector given by 
$$
\textbf{w}=[w_0, \cdots , w_{K+2}] \ .
$$
We also define an averaging operator 
\begin{equation}
\label{eq:consensus_operator}
{\bf G} ( {\bf w})= \left[ \bar{w} , \cdots , \bar{w} \right] \ ,
\end{equation}
where $\bar{w}$ is a weighted average of the input vector components given by
\begin{equation} \label{eq:bar_w}
\bar{w} = \sum_{k=0}^{K+2}\mu_k w_k,
\end{equation}
and $\mu_k$ is the weight for each agent, computed as
$$
\mu_k= 
\begin{cases}
    \frac{1}{K(1+\beta)},&  0 \leq k < K \\
    \frac{\beta}{3(1+\beta)},&  K \leq k < K+3\\
\end{cases} \ .
$$
Notice that $\beta$ then provides a mechanism to weight the amount of regularization relative to the data-fitting agents.

Using this notation, the MACE equilibrium equation is
\begin{equation}
\label{eq:MACE}
{\bf F} ( {\bf w})= {\bf G} ( {\bf w}) \ . 
\end{equation}
This equation enforces that all agents have the same output value (consensus) and that the vectors $\delta_j = w_j - F_j(w_j)$ satisfy $\delta_j = 0$ (equilibrium) \cite{CE}.  

\section{Computing the MACE Solution}

It is shown in \cite{CE} that the solution to (\ref{eq:MACE}) is also the fixed point of the operator ${\bf T}=(2{\bf G}-I)(2{\bf F}-I)$. 
One popular method of finding such a fixed point is Mann iteration
\begin{equation}
    {\bf w} \leftarrow(1-\rho) {\bf w}+\rho {\bf T} {\bf w},
\end{equation}
where $\rho\in (0,1)$ controls the convergence speed.

\begin{algorithm}
\caption{General MACE algorithm}\label{MACE_algorithm}
\textbf{Input}: Initial Reconstruction: $x^{(0)} \in \mathbb{R}^N$ \\
\textbf{Output}: Final Reconstruction: $x^* \in \mathbb{R}^N$
\begin{algorithmic}[1]
\State ${\bf w} \leftarrow[x^{(0)},...,x^{(0)}]$
\While {not converged}
\State ${\bf x} \leftarrow {\bf F} ( {\bf w} )$
\State ${\bf z} \leftarrow {\bf G} ( 2{\bf x} - {\bf w})$
\State ${\bf w} \leftarrow {\bf w} + 2\rho({\bf z}-{\bf x})$
\EndWhile
\State \Return $x^*\leftarrow\sum_{k=1}^{K+M}\mu_kx_k$
\end{algorithmic}
\end{algorithm}

Algorithm \ref{MACE_algorithm} shows the general method of solving MACE with Mann iterations.
The algorithm starts from an initial reconstruction $x^{(0)}$, and uses Mann iterations to find the equilibrium point between the prior and forward model terms. 
From \cite{CE}, when the agents $F_k$ and $H_m$ are all proximal maps of associated cost functions $f_k$ and $h_m$, this equilibrium point is exactly the solution to the consensus optimization problem of (\ref{eq:inverse_problem}).

\section{Experimental Results}
In this section, we present experimental results derived from an actual CT scan of a pipettor device, from which we simulated a 3D dataset with two poses.

\begin{figure*}[t]
\centering
   \includegraphics[width=\textwidth]{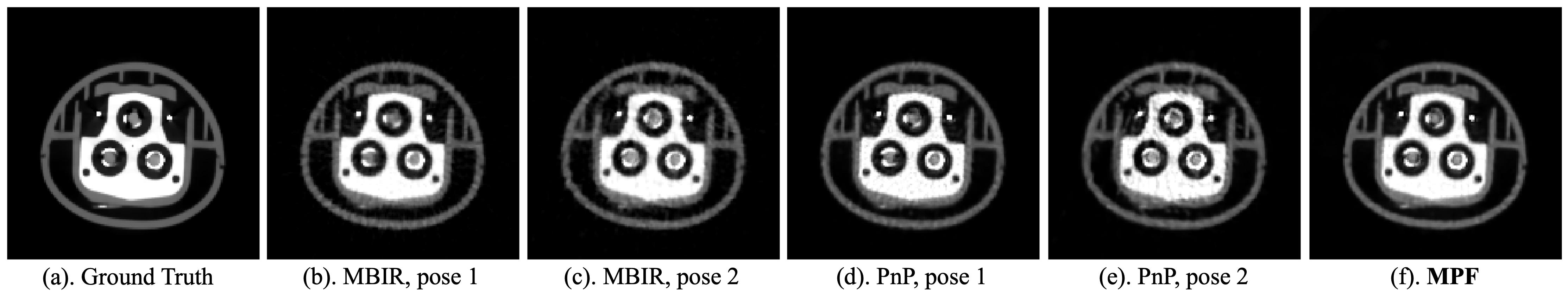}
   \caption{Comparison of different reconstruction methods and different poses in XY plane. Multi-Pose Fusion reduces sparse-view artifacts compared to single pose reconstructions.}
   \label{fig:results_axial}
\end{figure*}

\begin{figure*}[t]
\centering
   \includegraphics[width=\textwidth]{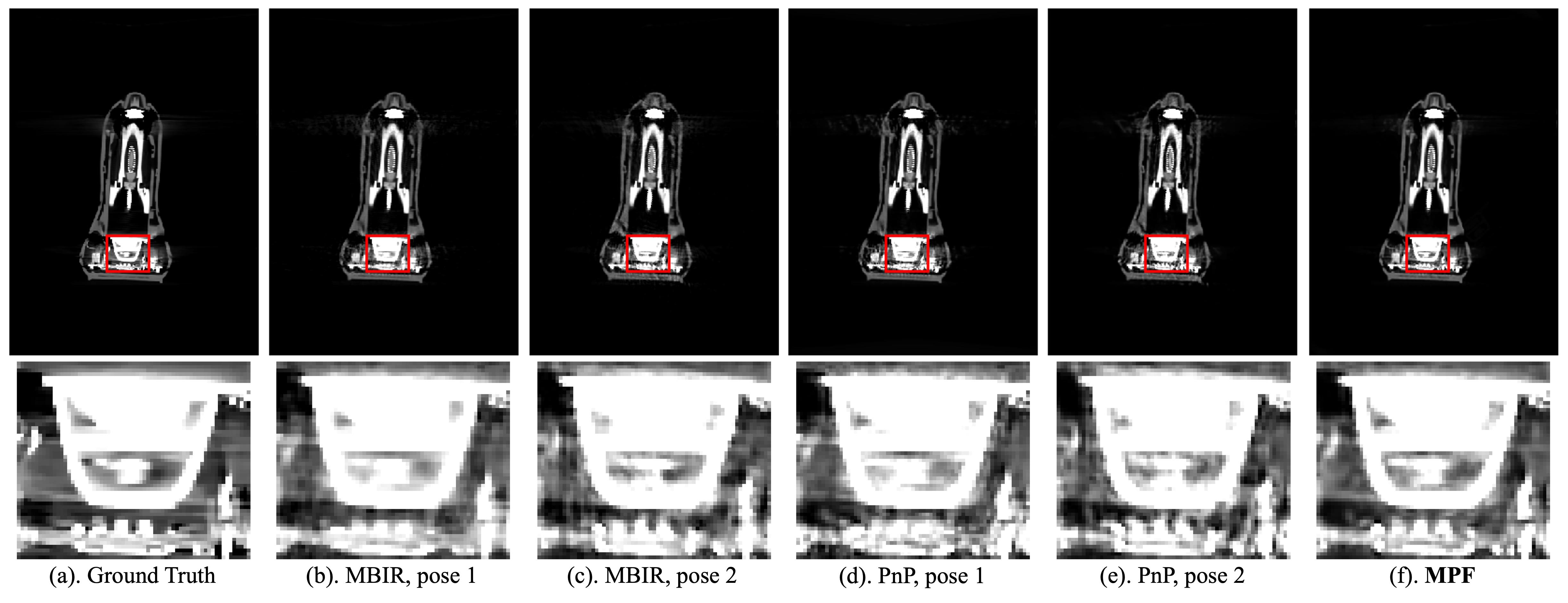}
   \caption{Comparison of different reconstruction methods and different poses in YZ plane. The top row shows a coronal slice for the entire region of reconstruction, and the bottom row shows a local region in the coronal slice marked with red box. MPF recovers fine details of the object that are either missing or poorly reconstructed in single pose reconstructions}
   \label{fig:results_coronal}
\end{figure*}

The ground truth image was generated from an MBIR reconstruction of a pipettor scan. The CT scan was acquired on a cone-beam system with 2100 views spanning a $360^{\circ}$ view angle, and the detector size was $480$ rows $\times$ $384$ channels. The MBIR reconstruction used a qGGMRF prior to incorporate regularization \cite{qGGMRF}. An axial and a coronal slice of the ground truth is shown in Figure~\ref{fig:results_axial}(a) and \ref{fig:results_coronal}(a) respectively.

We first generate two poses of the ground truth image. The first pose is the original pose of the object, in which case the transformation map is the identity. The second pose is obtained by rotating the object by $45^{\circ}$ along the XZ plane and then by $30^{\circ}$ along the YZ plane. For this part of the simulation, the rotation is performed using the \emph{SciPy} library \cite{2020SciPy-NMeth} with an order-5 spline interpolation method. 

Next, we generate a synthetic cone-beam sinogram for each pose by using the forward projector of \cite{mbircone}. For each synthetic sinogram, we use 35 views spanning a $360^{\circ}$ view angle. 

Finally, we perform multi-pose fusion with Algorithm~\ref{MACE_algorithm} with agents from \eqref{eq:CGPM_final} and multi-slice fusion.   
The image transformations are implemented using an order-3 B-Spline interpolation method \cite{de1978practical}, and the transformation parameters are estimated to sub-pixel accuracy from two initial PnP reconstructions of each pose using the SimpleITK library \cite{simpleITK}.

We compare the proposed MPF algorithm with single pose reconstructions with MBIR and PnP in Figures~\ref{fig:results_axial} and \ref{fig:results_coronal}. 
The single pose PnP algorithm uses the multi-slice fusion approach in \cite{MACE4D} to form a 3D denoiser from three domain-specific DnCNN \cite{DnCNN} 2D denoisers.  For a fair comparison, the denoisers in PnP are identical to the denoisers in our proposed MPF algorithm. 

From Figure~\ref{fig:results_axial}, we notice that MPF is able to reduce sparse-view artifacts observed in all single pose reconstruction results. From Figure~\ref{fig:results_coronal}, we notice that MPF is able to recover fine details of the object that are either missing or poorly reconstructed in single pose reconstruction results. For example, MPF successfully recovers the two triangular holes at the top of the zoomed region, while they are poorly recovered in all single-pose reconstructions.

Table~\ref{tab:NRMSE} lists the normalized root mean squared error (NRMSE) of the entire 3D volume for different reconstruction results. 
Our proposed MPF algorithm outperforms all single pose algorithms by a large margin in terms of NRMSE. 
We notice that for both single pose algorithms, NRMSE of the first pose is smaller than the NRMSE of the second pose. 
This is most likely due to the non-exact inverse transformation of the second pose. 
For the first pose, where the transformation function is the identity, there is no mismatch between the transformation and its inverse, while for the second pose, the interpolated inverse is not exact. 
We also notice the anomalous result that for the single pose reconstructions, MBIR reconstructions are better than PnP reconstructions for both poses both visually and numerically. 
This warrants further investigation and may result from prior model mismatch given that the original ground-truth reconstructions were done with a qGGMRF prior.

\begin{table}
\caption{\label{tab:NRMSE}NRMSE of different reconstruction algorithms}
\begin{center}
\begin{tabularx}{0.45\textwidth} { 
  | >{\centering\arraybackslash}X 
  | >{\centering\arraybackslash}X | }
 \hline
 Method & NRMSE \\ \hhline{|=|=|}
MBIR, Pose 1 & 0.1454 \\
\hline
MBIR, Pose 2 & 0.1690 \\
\hline
PnP, Pose 1 & 0.1687 \\
\hline
PnP, Pose 2 & 0.1853 \\
\hline
\textbf{MPF}  & \textbf{0.1288} \\
\hline
\end{tabularx}
\end{center}
\end{table}

\section{Conclusion}

In this paper, we introduced Multi-Pose Fusion, a novel algorithm that performs joint tomographic reconstruction from multiple poses of a single object. Our method incorporates data fidelity from multiple poses into the MACE framework using Conjugate Proximal Maps. Our algorithm can be easily implemented with standard CT reconstruction software and an off-the-shelf image registration toolbox. Compared to the single pose reconstructions, Multi-Pose Fusion produces better reconstruction results than single-pose algorithms by reducing artifacts and recovering more details of the object.


\section*{Acknowledgment}
The research was supported by Eli Lilly and Company, NSF grant number CCF-1763896, and Oak Ridge National Laboratory.

\bibliographystyle{IEEEtran}
\bibliography{allerton.bib}

\end{document}